\documentclass[conference,compsoc]{IEEEtran}

\usepackage[nocompress]{cite}
\usepackage[pdftex]{graphicx}
\usepackage{tikz}
\usepackage{textcomp}
\usepackage{hyperref}

\newcommand\copyrighttext{%
  \footnotesize \textcopyright 2016 IEEE. Personal use of this material is permitted.
  Permission from IEEE must be obtained for all other uses, in any current or future
  media, including reprinting/republishing this material for advertising or promotional
  purposes, creating new collective works, for resale or redistribution to servers or
  lists, or reuse of any copyrighted component of this work in other works.
  DOI: \href{http://dx.doi.org/10.1109/ISSREW.2016.52}{10.1109/ISSREW.2016.52}
  }
\newcommand\copyrightnotice{%
\begin{tikzpicture}[remember picture,overlay]
\node[anchor=south,yshift=10pt] at (current page.south) {\fbox{\parbox{\dimexpr\textwidth-\fboxsep-\fboxrule\relax}{\copyrighttext}}};
\end{tikzpicture}%
}

\begin{document}

\title{A Platform for Automating Chaos Experiments}

\author{\IEEEauthorblockN{Ali Basiri,
Aaron Blohowiak,
Lorin Hochstein and
Casey Rosenthal}
\IEEEauthorblockA{Netflix\\
    \{abasiri,ablohowiak,lhochstein,crosenthal\}@netflix.com
}}



\maketitle
\copyrightnotice

\begin{abstract}
The Netflix video streaming system is composed of many interacting services. In
such a large system, failures in individual services are not uncommon. This
paper describes the Chaos Automation Platform, a system for running failure
injection experiments on the production system to verify that failures in
non-critical services do not result in system outages.
\end{abstract}


\section{Introduction}
To an end-user, Netflix is a single service that allows them to stream television
shows and movies over the Internet. To the engineers who work for the company,
Netflix is a distributed system made up of many services that interact via
remote procedure call (RPC), sometimes referred to as a microservice
architecture \cite{microservices}.

In a large system such as Netflix, where hundreds of services run on thousands of
machines and engineers are making changes every day, many things can go wrong.
Fortunately, many of the internal services that make up Netflix are not critical
for the user to be able to watch a video. For example, a personalized list of
recommendations and bookmarks that recall where you left off when previously
watching a video add value to the user, but if the services that implement these
features stop working, we should still be able to provide a reasonable user
experience. Hodges describes this kind of graceful degradation as \textit{partial
availability} \cite{youngbloods}.

Partial availability doesn't come for free: engineers must explicitly implement
fallback behavior when making RPC calls against non-critical services. If
fallback behavior is not implemented correctly, a problem in a non-critical
service can lead to an outage. This work addresses the following question: \textit{how
can we have confidence that Netflix users will still be able to stream videos
after non-critical services have failed}? 

At Netflix, we practice Chaos Engineering \cite{chaoseng}. 
Namely, we believe there is a level of complexity in modern distributed systems that is
chaotic, and that a chief architect cannot hold all of the system's moving parts
in their head. Chaos Engineering is about engineering practices that help
us surface systemic effects, as embodied by the Principles of Chaos
Engineering \cite{PoC}.

In particular, we believe that to have maximum confidence you must test in your
production environment with live traffic.  Chaos Monkey \cite{chaosmonkey} is one
example of Chaos Engineering in practice at Netflix.  Another example is
automated canary analysis \cite{canary}, which tests new code in the production
environment with live traffic.  Unfortunately, canary analysis is not guaranteed
to test the code paths associated with dealing with failures in non-critical
services.  Another tenet of Chaos Engineering is automation: we want an
automated solution for ensuring the system is resilient to failures in
non-critical services.

This paper describes our proposed solution: the Chaos Automation Platform, or
ChAP. ChAP enables engineering teams to run Chaos Engineering experiments on
live traffic in production in order to build confidence that their service will
degrade gracefully when non-critical downstream services fail.

ChAP works by diverting a fraction of production traffic, injecting failures
into the diverted traffic, and checking that the system behaves as expected.
Section~\ref{sec-chap-expt} describes how an engineer would use ChAP to verify that
Netflix is resilient to failures in a particular service.

\section{Individual service failures vs system-level failures} \label{sec-failure}

As Hodges notes, ``distributed systems are different because they fail
often'' \cite{youngbloods}. When a system runs on thousands of servers, it becomes
very likely that something will go wrong somewhere.

A simple example of a failure is a bug that results in an
unhandled exception\footnote{At Netflix, most services are implemented in Java,
which uses exceptions for error signaling.}, such as a null pointer exception.
In Netflix's microservice architecture, an unhandled exception results in a
service returning an HTTP 500 error code \cite{rfc7231}.

There are other failure modes that are common for an individual service in a
microservice architecture. One common problem is resource exhaustion. Examples of
finite resources on a server include memory, disk space, CPU cycles, threads, and open TCP/IP
connections. When a server runs out of one of these resources, system calls that
would normally succeed may block or throw exceptions. Resource exhaustion can be
caused by a resource leak, but it may also occur if the load on a server exceeds
its capacity. Here the problem is that the service has been insufficiently
scaled: not enough servers have been allocated to that service.

When a server runs low on one of its resources, one symptom is an
increase in the average response time of the server. For example, memory
pressure on a server may lead to garbage collection pauses. Another
example: for a service that allocates one thread-per-request, if the number of
pending requests exceeds the number of available threads, latency will increase. 

Yet another issue is the environment that these services run in. All of the
Netflix services run within the Amazon Web Services Elastic Compute Cloud (EC2),
an infrastructure-as-a-service cloud computing environment \cite{nist}. Because
cloud providers such as EC2 compete on price, in order to reduce costs they use
commodity-grade hardware instead of more reliable enterprise-grade hardware.
This increases the likelihood of an individual server failing because of
hardware issues. Transient networking issues such as latency spikes are also not
uncommon in cloud computing environments. When deploying to a cloud computing
platform, it is the responsibility of the software engineers to design systems
that incorporate redundancy to compensate for occasional failures in hardware. 

Individual service failures are inevitable, and Netflix engineers 
leverage the Hystrix \cite{hystrix} library to implement fallback logic to handle
failures in downstream services.  Our goal is to prevent system-level failures.
In particular, our goal is to reduce the likelihood of an outage, when Netflix
customers are not able to stream videos. The primary metric of system health at
Netflix is the number of video stream starts per second, internally referred to
as SPS \cite{SPS}.

A failure of an individual service can lead to a drop in SPS if the client calling
the service does not have proper fallbacks in place. A study by Yuan et
al.\ revealed that 92\% of catastrophic system failures happened because of
incorrect error-handling logic \cite{Yuan14}.

\begin{figure}
\includegraphics[width=\columnwidth]{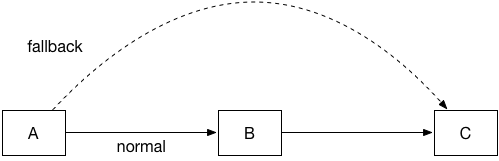}
\caption{Unexpected fallback behavior}
\label{fig-cascade}
\end{figure}

Even if fallback logic is present in a client, the failure of a non-critical
service may still lead to a system-level failure due to cascading effects.
Consider the following failure scenario, illustrated in Figure~\ref{fig-cascade}.

Typically, service \textit{A} calls service \textit{B}. For some reason,
\textit{C} starts to become overloaded, and returns errors to \textit{B}. The
fallback behavior for \textit{B} is not working correctly, which causes
\textit{B} to return errors. \textit{A} detects a problem and calls \textit{C}
as a fallback.  Fallback behavior that should have alleviated the load on
\textit{C} instead increased the load on \textit{C}, accelerating the problem
and resulting in an outage.

We believe that ChAP will help us identify these kinds of failure
modes before they result in outages.

\section{Example of a non-critical service: gallery} \label{sec-galleries}

When a user logs in to Netflix, they are presented with rows of images, called
\textit{galleries}, that represent video content. Each gallery represents a different
category. Examples of galleries include:

\begin{itemize}
    \item Trending Now
    \item Recently Added
    \item Critically-acclaimed Comedies
    \item TV Dramas
\end{itemize}

The list of galleries and the contents of the gallery are personalized for each
Netflix user: different users will be shown different galleries.

The \textit{Gallery} microservice is responsible for generating the
galleries. If this service stops working, the client that calls the
Gallery service must return a sensible fallback. For example, it may return an
older gallery that is present in a local cache. Or, it may return
a gallery that is not personalized for the particular user. From the user's
perspective, the Netflix interface should still appear to be working properly,
even if the content presented to the user is stale or not fully personalized.

\begin{figure}
    \includegraphics[width=\columnwidth]{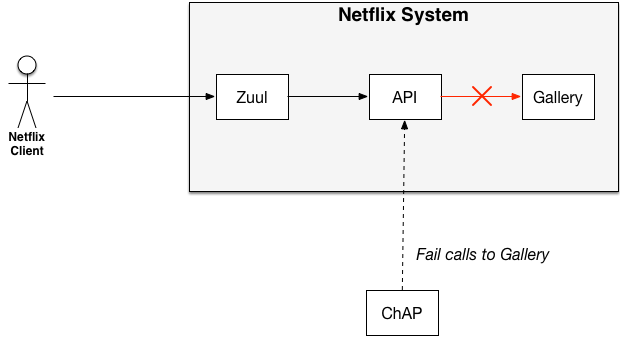}
    \caption{Services in the request path when calling Gallery}
    \label{fig-request-path}
\end{figure}

Figure~\ref{fig-request-path} shows the request path for requests that ultimately
reach the Gallery service. The first service in the request path is Zuul
\cite{zuul}, a reverse-proxy that serves as the front-door to Netflix. Next in the
request path is a service called API \cite{api}. API contains the Gallery client library
that makes calls against the Gallery service. It is this client library that is
responsible for serving fallbacks in the event that the Gallery service fails.
To verify that this fallback behavior works correctly, we must inject failures
on the calls from API to Gallery.

\section{Running a ChAP experiment} \label{sec-chap-expt}

Consider the following scenario: Alice, a (fictional) QA engineer on the Gallery team,
wants to verify that Netflix is resilient to failures in the Gallery service.
She uses ChAP's web interface to define an experiment. Because ChAP injects
failures on the client side of the request, she selects the API server group as
the subject of the experiment. She specifies that all calls to the Gallery
service should fail. She chooses to divert only a small amount of traffic for
this experiment: 0.3\%. She chooses a duration of 30 minutes for the experiment.

Finally, she selects the metrics that she is interested in observing for the
experiment. She chooses a number of Hystrix commands to track for the
experiment. Hystrix is a library that allows engineers to wrap RPC calls and
specify what the fallback behavior should be if an RPC call fails. Each Hystrix
command has a name, e.g.: ``GetGallery''.

For each Hystrix command, for the control and experiment server groups, ChAP will
display counts of:

\begin{itemize}
    \item successful requests served
    \item successful fallbacks served
    \item failed fallbacks served
\end{itemize}

\begin{figure*}
\includegraphics[width=\textwidth]{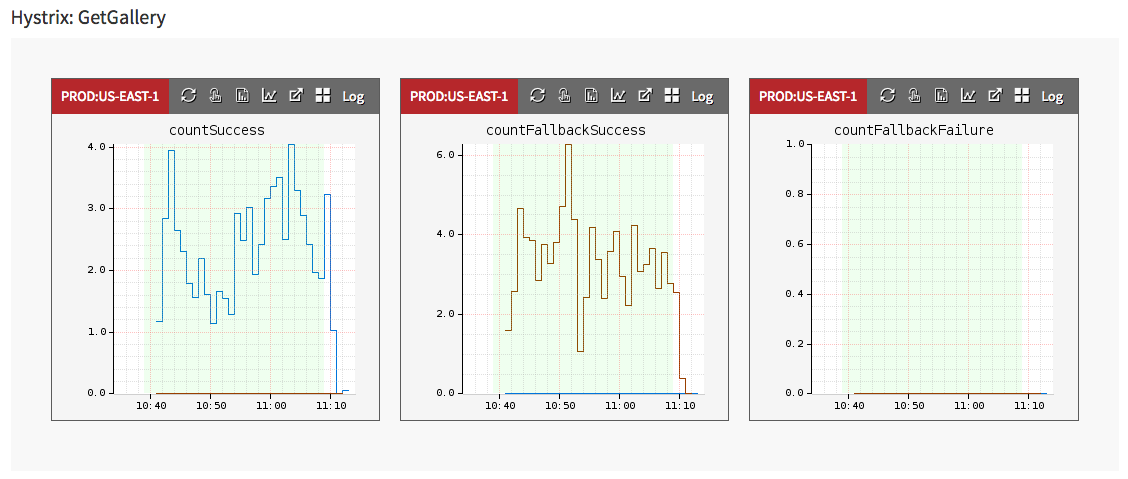}
\caption{ChAP plots for the GetGallery Hystrix command}
\label{fig-hystrix}
\end{figure*}

An example set of plots for the GetGallery Hystrix command is shown in Figure~\ref{fig-hystrix}.

Alice expects to see a large number of successful requests served in the
control group, and a large number of successful fallbacks served in the experiment
group.

\begin{figure*}
\includegraphics[width=\textwidth]{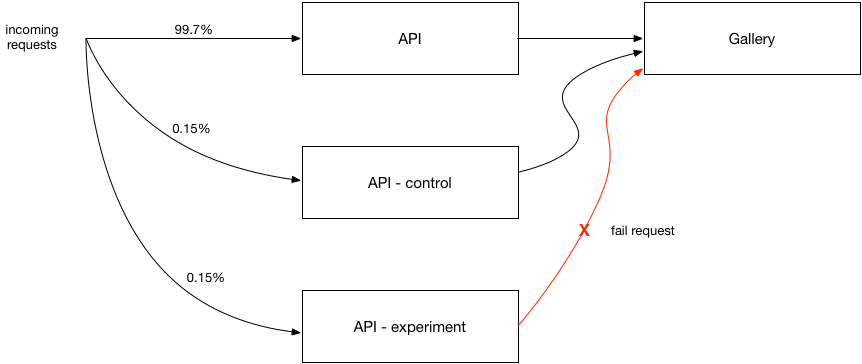}
\caption{A fraction of traffic is routed to the control and experiment server groups. Failures are only injected into the experiment server group.}
\label{fig-asgs}
\end{figure*}

Once the experiment starts, the following things happen, as depicted in Figure~\ref{fig-asgs}.

ChAP creates two new server groups, named api-chap-control and
api-chap-experiment. The servers in these two new groups are deployed
with the same software as the servers in the api server group.

Of all of the requests that are destined for the API services, 99.7\%
are routed to the original API server group, 0.15\% are routed to the
api-chap-control group, and 0.15\% are routed to the
api-chap-experiment group. In the api-chap-experiment group, all of the RPC
calls to the Gallery service fail immediately with an error.

ChAP presents Alice with a dashboard that plots the metrics specified by
user for the control and experiment groups. The dashboard also shows the SPS for
each group. By comparing the metrics between the two groups, Alice can determine
whether the system is handling Gallery failures correctly.

\section{Implementation details}

ChAP uses an internally developed system called FIT \cite{fit} to cause RPC
calls between microservices to fail. FIT is only able to inject two types of
failures: an error response and an increase in latency. However, from the point
of view of a client making a call to a service, a large number of problems that
can occur in an individual service manifest as either an error response or a response delay.
Hence, ChAP can model many types of real failures in individual services.

ChAP works by coordinating among many existing systems
inside of Netflix. In addition to FIT, ChAP interacts with Hystrix \cite{hystrix}
(fault tolerance), Spinnaker \cite{spinnaker} (deployment), Eureka \cite{eureka}
(service discovery), Zuul \cite{zuul} (reverse-proxy), Archaius \cite{archaius} (dynamic
configuration management), Ribbon \cite{ribbon} (interprocess communication), Atlas
\cite{atlas} (telemetry) and Mantis \cite{mantis} (stream processing).

\section{Current status and future work}

ChAP is still under heavy development, with a few teams inside of Netflix
currently test-driving the system and providing feedback. Our ultimate goal is
to be able to detect automatically whether a service is resilient to failure
rather than relying on a human looking at dashboards and making a judgment. We
also plan to integrate ChAP into the Spinnaker deployment system so that ChAP
experiments can be started automatically as part of the deployment process.

There are failures that FIT (and, hence, ChAP) cannot currently model. We
can only inject failures in the \textit{request path}, in requests that originate from
a Netflix client device. In particular, we cannot yet inject failures in calls
between services that are occur during the startup of a service.

Finally, while we use SPS as our health metric, what we are ultimately concerned
about is the user experience. In the future, we hope to use information from
client devices to get more accurate information on the impact of a ChAP
experiment on a user.

\bibliographystyle{IEEEtran}
\bibliography{refs}

\end{document}